# Experimental demonstration of longitudinal beam phase space linearizer in a free-electron laser facility by corrugated structures

Haixiao Deng,[1] Meng Zhang,[1] Chao Feng,[1] Tong Zhang,[1] Xingtao Wang,[1] Taihe Lan,[1] Lie Feng,[1] Wenyan Zhang,[1] Xiaoqing Liu,[1] Haifeng Yao,[1] Lei Shen,[1] Bin Li,[1] Junqiang Zhang,[1] Xuan Li,[1] Wencheng Fang,[1] Dan Wang,[2] Marie-emmanuelle Couprie,[3] Guoqiang Lin,[1] Bo Liu,[1] Qiang Gu,[1] Dong Wang,[1] and Zhentang Zhao[1]

[1] *Shanghai Institute of Applied Physics, Chinese Academy of Sciences, Shanghai, 201800, P. R. China*
[2] *Department of Engineering Physics, Tsinghua University, Beijing 100084, P. R. China*
[3] *Synchrotron SOLEIL, L' Orme des Merisiers, Saint-Aubin, BP 48, 91 192 Gif-sur-Yvette, France*

Removal of residual linear energy chirp and intrinsic nonlinear energy curvature in the relativistic electron beam from radiofrequency linear accelerator is of paramount importance for efficient lasing of a high-gain free-electron laser. Recently, it was theoretically and experimentally demonstrated that the longitudinal wakefield excited by the electrons itself in the corrugated structure allows for precise control of the electron beam phase space. In this Letter, we report the first utilization of a corrugated structure as beam linearizer in the operation of a seeded free-electron laser driven by a 140 MeV linear accelerator, where a gain of ~10,000 over spontaneous emission was achieved at the second harmonic of the 1047 nm seed laser, and a free-electron laser bandwidth narrowing by about 50% was observed, in good agreement with the theoretical expectations.

The recent advent of X-ray free-electron laser (FEL) presenting high-brightness, ultra-short pulse, tunable wavelength, and flexible polarization, pushes nonlinear optics science into the x-ray domain. FELs open new frontiers of ultrafast science in multidisciplinary investigation of matter and pave the way for various novel time-resolved experiments [1-2]. FEL generates coherent radiation by wiggling relativistic electron beam when passing through a transverse periodic magnetic field, i.e. the undulator, with the capability of tuning radiation frequencies from the infrared to hard X-ray regions [3-8]. In order to achieve efficient lasing for outstanding FEL performance, precise control of the electron beam phase space is required, mainly for high peak current and/or constant energy profile along the longitudinal dimension. In a linear accelerator (LINAC) driven high-gain FEL, the electron beam emitted from the cathode is usually accelerated in radiofrequency (RF) cavities and temporally compressed in magnetic chicanes, which leaves an undesired time-energy correlation in the electron bunch, i.e., linear energy chirp and nonlinear energy curvature [9-10]. Such time-energy correlations are usually detrimental to the FEL performance. As well illustrated theoretically and experimentally [11-15], with a linear energy chirp of $h$, the output wavelength of a seeded FEL, e.g., high-gain harmonic generation (HGHG) [16] will shift to

$$\lambda_{HGHG} = \lambda_s (1 + hR_{56})/n, \qquad (1)$$

where $\lambda_s$ is the seed wavelength, $R_{56}$ is the dispersion of the dispersive chicane, and $n$ is the harmonic number. Thus, a positive linear energy chirp leads a redshift while the high-order nonlinear one contributes to a bandwidth broadening. Although the degradations of FEL spectrum in a HGHG scheme due to the beam energy correlations can be partly cured by the recent advanced seeding concepts, e.g., echo-enabled harmonic generation [17-18] and phase-merging enhanced harmonic generation [19-21], it is preferred that the energy correlations are removed before FEL process.

Typically, the removal of linear energy chirp is done by cooperation of off-crest acceleration and wakefields induced in downstream accelerator cavities, while the correction of nonlinear curvature is accomplished in an active harmonic cavity (e.g., Ref. [6]). Another more sophisticated method uses sextupoles to control second-order nonlinear effects and thereby optimizes the compression and shaping of electron bunch in achromatic sections [22-23]. However, these solutions are costly and inefficient in the next generation FELs driven by the superconducting LINAC.

An alternative, RF-free approach to correct the undesired energy correlation in the electron beam is dedicated structure, which can intentionally generate strong longitudinal wakefield. The first structure suggested for such purposes is a beam pipe with a thin dielectric layer [24]. Similar proposals including a resistive pipe of small radius [25] and a metallic pipe with periodic corrugations [26] were put forward later. Currently, the corrugated metallic structure attracts great interest of accelerator and light source community, as its wakefield may reach the maximal strength for a given aperture. Moreover, both frequency and amplitude of the wake can be easily adjusted by choosing the corrugated structure parameters. Theoretical investigations indicate that, corrugated structures can be used to remove the residual linear energy chirp [26], to linearize the high-order RF curvatures [27-28], to stabilize the beam energy jitter [29] and to emit high power terahertz wave [30], provided that the electron bunch length and the structure parameters are well matched. Recently, experiments using passive wakefields to remove linear energy correlation of the electron beam have been successfully demonstrated with dielectric structures [31-32] and metallic corrugated devices [33-34]. Therefore, it is widely believed that such passive wakefield devices can significantly reduce cost and improve FEL performance, and almost every new FEL facility is seriously considering the idea of using corrugated structures to control the longitudinal phase space of the electron beam [35-38].

So far, the effect of a corrugated structure was just tested on LINACs as a beam dechirper [33-34]. In this Letter, we report the demonstration of a metallic corrugated structure serving as a beam linearizer in a seeded FEL. An obvious FEL bandwidth narrowing by 50% and a FEL central wavelength shift of 8nm in the experiment confirm the feasibility to employ a corrugated device for precise control of the electron beam, and thus to improve the output performances in future x-ray FEL user facilities.

Geometry of a corrugated structure is plotted in Fig. 1(a), which can be manufactured by low-speed wire-cut electrical discharge machining. Usually a rectangular geometry is preferred since it provides operational flexibility and can be effective for different electron bunch cases. The corrugations are characterized by period $p$, depth $\delta$, gap $g$, length $l$ and width $w$. According to wakefield theory [26, 35, 39-40], in the wide structure limit, the longitudinal wakefield generated by an electron passing the corrugated plates is a damped cosine oscillation, which can be numerically fitted as follows

$$w(z) = \left(\frac{\pi^2}{16}\right)\frac{Z_0 c}{\pi a^2} H(z) F e^{-\frac{kz}{2Q}} \cos(kz), \qquad (2)$$

where $Z_0=377\ \Omega$ is the impedance of free space, $H(z)$ is the step function, $c$ is the speed of light, $z$ is the coordinate along the beam path, and $F$, $Q$ and $k$ are the amplitude factor, wave number and quality factor, respectively, which are functions of corrugation parameters [40]. The dominant wave number of the wakefield can be approximated as $k = [p/(a\delta g)]^{1/2}$. The wakefield produced by a beam can be found by convolving the wake function with the longitudinal bunch profile. When the induced energy loss along the electron beam is a sinusoidal waveform with a wavelength of about twice as the bunch length, it can be used as a linearizer for the temporal phase space. More accurate 3D electric field excited by the electron bunch when passing through the corrugated device can be solved by CST simulation [41], as illustrated in Fig. 1(b), which is in good agreement with the theory.

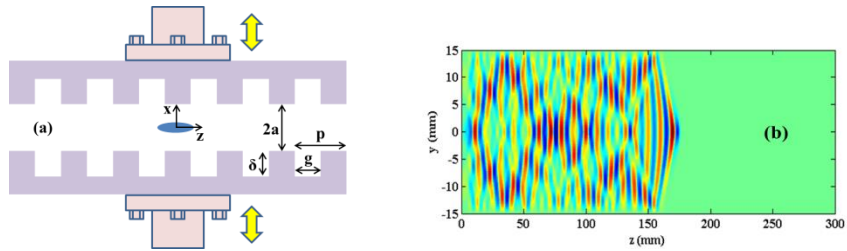

FIG. 1 (color online). (a) geometry of a moveable corrugated plates, where a rectangular coordinate system is centered on the axis. For the case demonstrated in this Letter, the separation between the two aluminum plates can be adjusted from $a$=1 mm to $a$=25 mm, with parameters $p$=0.6 mm, $\delta$=2.0 mm and $g$=0.3 mm, and plates length of 0.3 m and plates width of 30 mm. (b) 2D snapshot of excited electric field in the corrugated structure ($a$=1 mm), as calculated with 3D method, while an 8.8 ps (FWHM) electron beam is passing through the corrugated plates at $x$=0, $y$=0 and $z$=175 mm.

As shown in Fig. 2, the experiment was conducted in an HGHG-FEL lasing process at Shanghai deep ultraviolet FEL (SDUV-FEL) facility [42-45], which is a test bench for novel FEL principles and key technologies for X-ray FEL. The LINAC consists of an S-band photo-injector (150 pC bunch charge and 30 MeV beam energy), and four 3 m S-band accelerating modules (A1-A4) symmetrically placed on both sides of the magnetic compressor. The LINAC provides 140 MeV beams, with normalized emittance of 4 μm-rad, global energy spread of 0.1%, and pulse duration of 8.8 ps (FWHM) without compression. The seed of HGHG-FEL is an external 1047 nm laser with pulse length of 8.7 ps (FWHM) and tunable pulse energy up to 100 μJ. The beam energy modulation is achieved when the electron beam and laser beam overlap spatially and temporally in the short electromagnetic undulator (i.e., modulator after UNPRF0, 10 periods with period of 65 mm, and variable magnetic field) and, afterwards, the electron beam passes through a dispersive chicane with $R_{56}$ of 0-70 mm, to form the density modulation, where the phase/amplitude properties of the seed pulse are encoded into the electrons. Finally, the micro-bunched electron beam is sent through the radiator sections, which comprises two segments of 1.6 m long undulator with period of 40 mm, to produce coherent FEL radiation at the 2$^{nd}$ harmonic of the seed laser. Considering the requirement for the beam phase space linearization and the free space available at SDUV-FEL, a pair of corrugated plates wrapped in a 0.5 m vacuum chamber was equipped after the modulator. The chamber support includes two separate motors to allow for independent positioning for each corrugated jaw, providing remote control of the full horizontal separation with 4 μm resolution, as well as the horizontal central position.

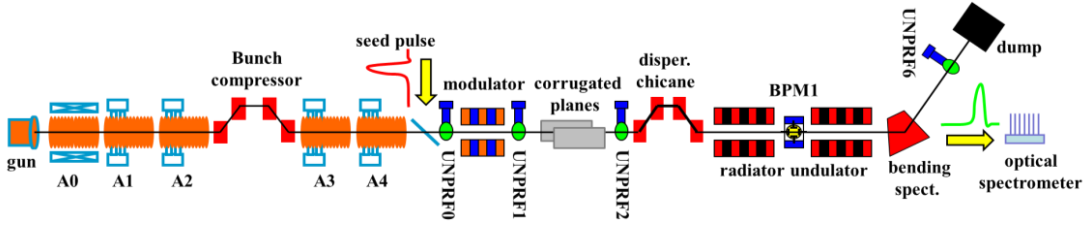

FIG. 2 (color online). Schematic layout of experiment at SDUV-FEL. In the experiment, several yttrium aluminum garnet (YAG) and optical transition radiation (OTR) screens are routinely used for the electron and laser beam position and size measurements. Two correctors attached at the two edges of the electromagnetic modulator are firstly fine-tuned for beam trajectory control in the corrugated structures, while the signals from screen UNPRF2 and BPM1 serve as the reference monitor of the bunch charge, thus to make sure that the electron beam passes through the corrugated chamber without too much loss during the whole experiment. Then the other two screens UNPRF0 and UNPRF1 are used to spatially synchronize the electron beam and laser beam in the modulator by adjusting the laser beam optical path. The FEL radiation properties were investigated with a spectrometer, a photo detector and a charge-coupled device (CCD), which were placed close to an in-vacuum reflecting mirror downstream of the radiator undulators.

For predicting the FEL performance dependence on the corrugated wakefield linearizer, we carried out self-consistent 3D simulations [46] in which all macro-particles are directly tracked from the photocathode to the undulator exit by using a set of well-benchmarked codes. Firstly, the macro-particles generated from a Monte Carlo sampling are tracked by ASTRA [47] with 6D phase space in the photo-injector. Then the macro-particles are further tracked by ELEGANT [48] for energy booster. The laser-beam interactions in the modulator are determined by an algorithm on the basis of the fundamentals of electrodynamics, i.e. the electron's behavior is determined by the magnetic field and the laser electric field in the time domain [49]. The pass through of corrugated chambers is implemented in ELEGANT, where the corrugated plate is replaced by a pipe with an equivalent longitudinal wakefield. After the particle dumping from ELEGANT, the whole bunch is loaded into GENESIS [50] to run a FEL simulation.

Three cases with the accelerating phase of A3 and A4 being 0°, -25° and 25° are considered here, as illustrated by the simulation results in Fig. 3. In the case of on-crest acceleration shown in Fig. 3(a), the projected beam energy spread can be significantly suppressed by a factor about 50%. While in two other off-crest acceleration cases shown in Fig. 3(b) and 3(c), the nonlinear beam energy correlations induced by the RF curvature were corrected by wakefield of the corrugated plates. It is worth stressing that, for a seeded FEL configuration under such a linearization, besides a FEL bandwidth reduction, a FEL central wavelength shift may also be induced, depending on the beam energy chirp variation at the core of the electron beam.

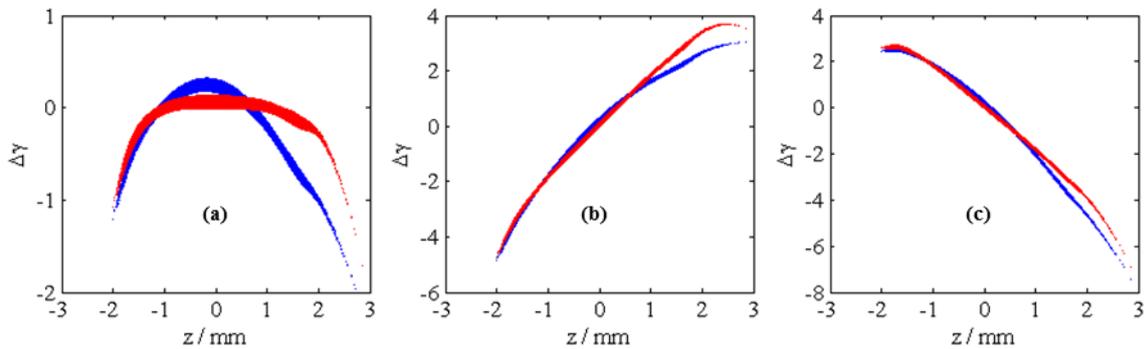

FIG. 3: (color online) Simulated phase space of the electron beam, where the blue and red represents the phase space before and after the corrugated device, respectively, with the phases of A3 and A4 being (a) 0°, (b) -25° and (c) 25°.

The corrugated structure experiments at the SDUV-FEL were performed at the 2$^{nd}$ harmonic of the laser seeding, and the accelerating phases of A3 and A4 were 25°; thus, beam energy correlations shown in Fig. 3(c) were expected before and after the linearization. In order to clearly distinguish the corrugated plates effects from the machine instabilities, a relatively large dispersion of $R_{56}$=6.5 mm was used. FEL simulations were carried out under the experimental conditions. As the simulated results shown in Fig. 4(a), a linear energy chirp of $h \approx 7.8$ m$^{-1}$ leads an output central wavelength shift from 523 nm to 550 nm for the case without beam linearization. The central wavelength is further shifted to 558 nm with an appropriate linearization. Simultaneously, the FEL bandwidth broadening due to the nonlinear RF curvature is significantly compensated by the wakefield. In the experiment, the seed laser was firstly optimized in the modulator to achieve maximum FEL radiation. At this stage, the two corrugated plates were opened ($a$=3.0mm) and a 550 nm HGHG signal with 7.8 nm bandwidth was observed (see the blue in Fig. 4(b)). Then the two corrugated plates were gradually closed to compensate the nonlinear RF curvature, which led to a wavelength shifting and bandwidth narrowing of HGHG output. When the corrugated plates were in its optimized gap ($a$ = 1.0 mm), the HGHG spectrum located at 558 nm (Fig. 4(b), red line) with a intensity relatively larger than that of before correction, and a spectral bandwidth of 3.7 nm. To show the repeatability of the measurements, eight consecutive radiation spectra observed from experiment are illustrated in Fig. 4(c). The HGHG bandwidth without beam linearization, averaged over eight shots, is found to be about 7.34±0.47 nm while that for appropriate linearization case is only about 3.75±0.26 nm. The averaged spectral brightness is found to be about 50% higher with corrugated linearizer. The experimental data agree well with the simulation results.

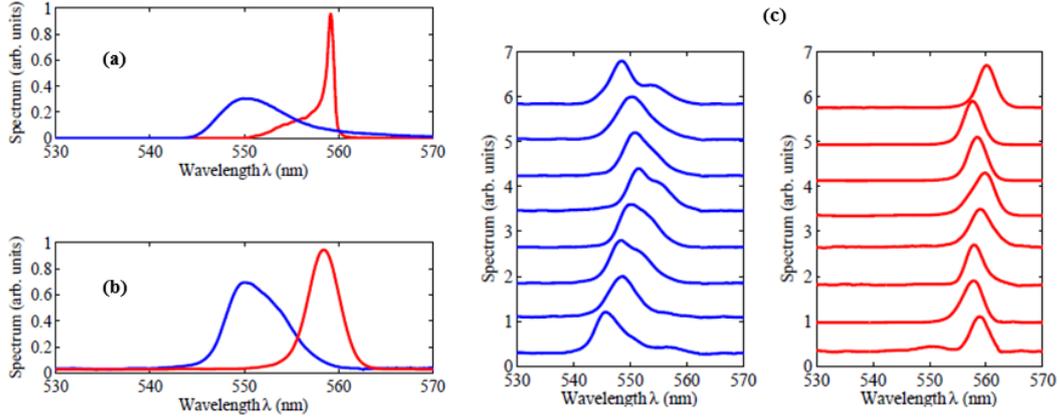

FIG. 4: (color online) FEL spectra from the simulation and experiment for two corrugated plane gaps (red @ $a$=1.0 mm and blue @ $a$=3.0 mm). (a) simulated radiation spectra; (b) typical spectra from experiment. the measurement were performed with a commercial spectrometer (model TRIAX-550, Jobin Yvon) with a focal length of 550 mm, a grating of 600 lines/mm, and an estimated resolution of 2.7 nm with the front slit open; (c) sequence of 8 consecutive shots FEL spectra from experiment.

The pulse energy was measured with SXUV100 photo-detectors (International Radiation Detectors) and home-made amplifiers. The efficiency was calibrated with a pulsed laser with an uncertainty of 30%. The measured pulse energy of the HGHG-FEL is about 200 nJ during the experiment, in reasonable agreement with numerical simulations, which means that a gain larger than $1 \times 10^4$ is achieved compared with the spontaneous radiation from one segment radiator. One may also note that, the FEL spectrum would be amplified and purified further during the exponential growth process in the radiator, thus in order to illustrate the electron beam energy correlation effect, FEL lasing here was accomplished without any beam compression, i.e., with a much lower FEL gain.

Moreover, a 30 °bending spectrometer line with screen UNPRF6 (a 300-kpx CCD camera with 7.5 μm/pixel resolution) located downstream of the undulators is used for beam energy and global beam energy spread measurement. Under on-crest acceleration case, a selection of beam energy spectrum measurements at different plate separations was carried out to verify the energy spread suppression by wakefield compensation of the corrugated plates. Fig. 5 shows the measured energy distribution of the electrons with various wakefield strengths. The relative electron beam energy spread was suppressed from $1.1 \times 10^{-3}$ to $7.5 \times 10^{-4}$, which is quite consistent with the beam tracking results, if one considers the contribution of the horizontal beam size.

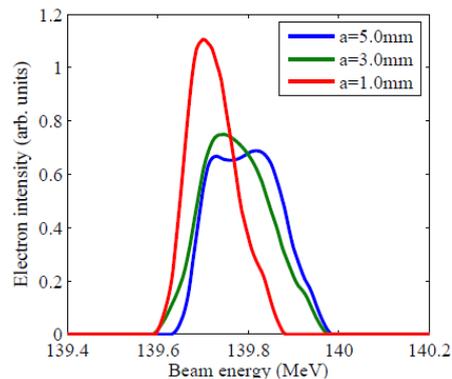

FIG. 5: (color online) The beam energy distribution measurement with on-crest acceleration of A3 and A4 cavities, in which the corrugated linearizer with separation parameter *a* = 5.0, 3.0 and 1.0 mm, respectively.

In summary, we experimentally demonstrated the possibility to linearize the beam energy curvature imprinted by the RF field during the beam acceleration, by using a twin-plane with metallic corrugations, in a laser seeding HGHG-FEL configuration at the Shanghai deep ultraviolet FEL test facility. By closing the corrugated linearizer separation from 6.0 to 2.0 mm, the initial FEL bandwidth of 7.34 nm was gradually reduced to 3.75 nm, together with an approximate 8.0 nm redshift of the FEL central wavelength. The residual bandwidth corresponded to the even higher order RF curvature, the imperfection of the 1047 nm seed laser, not fully optimized experimental condition, and mainly the limited spectrometer resolution of 2.7 nm. Reasonably good agreement between the measurement and model are observed, and our results give one confidence in the practicality of including a corrugated-plane linearizer in the design of next-generation FELs. As theoretically and experimentally predicted by early reports, when properly scaled, the corrugated devices can be used to remove the residual correlated energy spread at the end of the LINACs, and improve the beam stability due to on-crest acceleration after the bunch compressor in a low energy machine. Therefore, the corrugated structure can significantly simplify LINAC design and improve the performance of FELs, especially in seeded FELs in pursuit of fully coherence.

This technique can be easily extended to compensate the electron beam energy spread in the advanced single-shot MeV transmission electron microscopy [51-52], where a low energy spread is very critical to minimize the effects of spherical and chromatic aberrations and reach spatial resolution down to 10 nm. Current concept of the beam energy compensator relies on precise and high stability control of the amplitudes and phases of the S-band and X-band RF sources, requiring the recent advances on RF-laser synchronization and high stability modulator technologies. As we have demonstrated here, if one considers the pretty small electron beam size in a transmission electron microscopy, a tunable corrugated linearizer is a promising choice for beam energy spread compensation, and it is expected that no additional instabilities will be induced in such a passive linearizer.

*The authors thank SDUV-FEL operation staff for excellent support during the experiment. Thanks also go to D. Xiang, Z. Huang and G. Stupakov for their helpful comments and useful discussions. This work was partially supported by the Major State Basic Research Development Program of China (2011CB808300) and the Natural Science Foundation of China (11175240, 11205234, 11322550 and 11475250).*